\documentclass[%
aip,
amsmath,amssymb,
reprint, twocolumn%
,groupedaddress,nobibnotes,nofootinbib]{revtex4}
\usepackage{scrextend}
\usepackage{graphicx}% Include figure files
\usepackage{dcolumn}% Align table columns on decimal point
\usepackage{bm}% bold math
\usepackage{xcolor}
\usepackage[utf8]{inputenc}
\usepackage{courier}
\usepackage{amsmath}
\usepackage{etoolbox}
\usepackage{graphicx}
\usepackage[switch]{lineno}
\setcitestyle{super}
\renewcommand{\thefootnote}{\textcolor{black}{\@A}}
\DeclareMathAlphabet{\altmathcal}{OMS}{cmsy}{m}{n}
\makeatletter
\def\@email#1#2{%
	\endgroup
	\patchcmd{\titleblock@produce}
	{\frontmatter@RRAPformat}
	{\frontmatter@RRAPformat{\produce@RRAP{*#1\href{mailto:#2}{#2}}}\frontmatter@RRAPformat}
	{}{}
}%
\makeatother

\begin{document}
	
	\preprint{AIP/123-QED}
	
	\title{\Large  Gaussian pseudo-Orthogonal Ensemble of Real Random Matrices}
	\author{ \large Sachin Kumar$^\dagger$, Amit Kumar, S M Yusuf$^\star$}
	%	\email{$^\dagger$sachinv@barc.gov.in, $^\star$smyusuf@barc.gov.in}
	\email{sachinv@barc.gov.in, smyusuf@barc.gov.in}
	\affiliation{Solid State Physics Division, Bhabha Atomic Research Centre, Mumbai, India}
	\affiliation{Homi Bhabha National Institute, Anushaktinagar, Mumbai 400094, India}
	\date{\today}
	\begin{abstract}
		\large
		% \linenumbers
		Here, using two real non-zero parameters $\lambda$ and $\mu$, we construct Gaussian pseudo-orthogonal ensembles of a large number $N$ of $n \times n$ ($n$ even and large) real pseudo-symmetric matrices under the metric $\eta$ using  $ \altmathcal {N}=n(n+1)/2$ elements independently drawn from a Gaussian random population and investigate the statistical properties of the eigenvalues. When $\lambda \mu >0$, we show that the pseudo-symmetric matrix is similar to a real symmetric matrix, consequently, all the eigenvalues are real and so the spectral distributions satisfy Wigner's statistics. But when $\lambda \mu <0$ the eigenvalues are either real or complex conjugate pairs. We find that these real eigenvalues exhibit intermediate statistics. We show that the diagonalizing matrices ${ \cal D}$ of these pseudo-symmetric matrices are pseudo-orthogonal under a constant metric $\zeta$ as $ \altmathcal{D}^t \zeta \altmathcal{D}= \zeta$,  and hence they belong to a pseudo-orthogonal group. These pseudo-symmetric matrices serve to represent the parity-time (PT)-symmetric quantum systems having exact (un-broken) or broken PT-symmetry.
	\end{abstract}
	\maketitle 
	\section{\large \label{sec:level1} Introduction}
	%\linenumbers
	\large
	Eigenvalues of a Hamiltonian of a physical system can be interpreted as eigenvalues of the matrix which is obtained in a complete orthonormal basis for the corresponding system.  In Random Matrix Theory (RMT)\cite{1,2,3}, the invariance properties of the complex many-body Hamiltonian are seen in a class of matrices and the spectral properties of the complex many-body Hamiltonian are then predicted thereof.$~$RMT has been widely used in the analysis of spectra of various physical systems, such as strongly correlated systems\cite{4}, quantum spin chains\cite{5}, and disordered quantum systems\cite{6}.$~$RMT has been applied in the investigation of space-domain reactor-noise problems to calculate the probability distribution of reactivities\cite{7}. Moreover, random matrix theory has been a natural tool for quantum information theory\cite{8} where the entanglement spectrum statistics of many-body quantum systems have been investigated in the framework of RMT. 
	%%%%%%%%%%%%%%%%%%%
	The nearest level spacing $\delta \epsilon = |\epsilon_{n+1}-\epsilon_{n}|$ distribution (NLSD) of ensemble of a $2 \times 2$ real symmetric matrices is well known as \cite{1} $p_W(s)=\frac {\pi s}{2}   \exp{ -\frac{\pi s^2}{4}}, s=\delta \epsilon/{<}\delta\epsilon{>}$;  where matrix elements $a,b,c$  are independently drawn from a Gaussian Probability Distribution Function (PDF). This is called the NLSD of Gaussian Orthogonal Ensemble (GOE) due to the orthogonal symmetry of real symmetric matrices. Wigner  surmised that even when $n$ becomes large ($n>>2)$, NLSD $p(s)$ remains approximately close to $p_W(s)$\cite{2}. The NLSD $p_W(s)$ is known as the Wigner distribution function. RMT was first introduced in statistics by Wishart and later applied to nuclear physics by Wigner, where the Wigner surmise $p_W(s)$ represents the spectral distribution of neutron-nucleus scattering resonances\cite{3}, and NLSD $p(s)$ of nuclear levels of the same angular momentum $J$ and parity $\pi$ display the Wigner's surmise $p_W(s)$\cite{9}, whereas the mixed levels show the Poisson statistics $p_P(s)=\exp{(-s)}$. In the case of quantum spin chains,\cite{5}, if the Hamiltonian is integrable by Bethe ansatz, NLSD $p(s)$ is given by Poisson distribution $p_P(s)$, and in case of non-integrable by Bethe ansatz, NLSD $p(s)$ is given by Wigner distribution $p_W(s)$. In Anderson model of disordered systems \cite{6}, which undergoes a phase transition between an insulating and a metallic phase as a function of the disorder strength (Anderson metal-insulator transition), in the insulating phase, the eigen-energies are Poisson distributed $p_P(s)$, and the metallic phase leads to a Wigner distribution $p_W(s)$ of the energy levels, but at the critical point between the two phases, an intermediate statistics $p(s)$, which describes most closely both the Wigner's distribution $p_W(s)$ (linear repulsion) at small spacings and the Poisson distribution (exponential tail) at large spacings, occurs. Random matrix models to describe such intermediate statistics\cite{10} have been proposed. Across the many-body localization transition \cite{11}, intermediate statistics interpolating between $p_W(s)$ and $p_P(s)$ is proposed to be  $p_{MBL}(s){=}C_1 s ^\beta e^{-C_2  s^{2-\gamma_P}}, \gamma_P{\le}1$, which has been referred as sub-Wigner statistics \cite{12}.$~$Moreover, topological transitions in a Josephson junction are described by the semi-Poisson distribution\cite{13} $p_{SP}(s)=4 s e^{-2s}$, which is a simpler form of intermediate level spacing distribution $p_{MBL}(s)$ in the limit $\gamma_P \rightarrow 1$.$~$The intermediate spectral statistics have also been found to occur in several other systems, such as pseudo-integrable billiards \cite{14} and quantum maps \cite{15}, molecular resonances in Er isotopes\cite{16}. In\cite{17}, statistical properties of structured random matrices present the intermediate statistics and it is argued to be more ubiquitous and universal than was considered so far in RMT.
	
	Our motivation stems from conjecture\cite{18} that a non-Hermitian complex PT(parity-time)-symmetric Hamiltonians, which have been associated with pseudo-Hermitian Hamiltonians\cite{19}, connected to their adjoints by a similarity transformation$~$given as $\eta H \eta^{-1}{=}H^\dagger$ under a generalized parity $\eta$, may also exhibit a set of isolated real eigenvalues. Specifically, the eigenvalues are expected to be either entirely real or occur in complex conjugate pairs, depending on whether a real parameter in the potential lies below or above a critical value\cite{18}.$ $Random matrix theory of PT-symmetric or pseudo-Hermitian quantum systems has been subjected to great interest of research due to a remarkable surge of interest in PT-symmetric quantum systems and has been investigated extensively in recent years \cite{20,21,22,23,24,25,26,27,28,29}. Initially, the ensembles proposed were restricted to the case of 2 × 2 pseudo-Hermitian matrices \cite{20}, and pseudo-Hermitian random matrix models were approached in the $N \times N$ case a few years later \cite{21,22,23},$~$and further,$~$a general formalism for pseudo-Hermitian random matrix models has been laid down in \cite{25}.$~$The level-spacing distribution of the pseudo-Hermitian Dicke model near the integrable limit is close to the Poisson distribution, while it is Wigner distribution for the ranges of the parameters for which the Hamiltonian is nonintegrable \cite{26}.$~$In Marinello\cite{27}{\it et al} investigated the statistical properties of eigenvalues of pseudo-Hermitian random matrices to find that spectrum splits into separated sets of real and complex conjugate eigenvalues, the real ones show characteristics of an intermediate incomplete spectrum, and on the other hand, the complex ones show repulsion compatible with cubic-order repulsion. Concerning pseudo-Hermitian random matrices, the collection of work by Pato{\it et al.}\cite{28} is worth mentioning. Moreover, a recent study \cite{29} on level statistics of real eigenvalues in non-Hermitian systems serves as effective tools for detecting quantum chaos, many-body localization, and real-complex transitions in non-Hermitian systems with symmetries. Application of non-Hermitian random matrices can be found in other areas of physics, such as in QCD at finite chemical potential, where the Dirac operator becomes non-Hermitian\cite{30}, as well as in lattice corrections\cite{31}. Furthermore, the distribution of real and complex eigenvalues of the non-Hermitian Wilson-Dirac operator has been obtained\cite{32}. In quantum chaos, the transition between closed and open systems has been modeled in Ref.[33]. As mentioned above, while several studies have explored pseudo-Hermitian random matrices, the behavior of pseudo-symmetric random matrices displaying Wigner's distribution and intermediate statistics, and capturing the crossover between these two regimes, has not been studied in detail. The present work expands on this by introducing ensembles within a unified framework that describes both the statistics and the crossover between them across the full parameter space.
	
	\indent Real non-symmetric matrices $H_{n\times n}$ may have both real and complex conjugate eigenvalues.  It can be shown that any real square matrix which is diagonalizable is pseudo-symmetric\cite{19} $\eta^{-1} H \eta{=}H^t$ under the metric $\eta{=}(DD^t)^{-1}$ or some other secular metric (constant matrix). Here $D$ is the diagonalizing matrix of $H$, and $t$ is the transpose operation. The number of real eigenvalues of a pseudo-symmetric matrix is influenced by the signature of the metric $\eta$, which reflects the number of positive and negative entries. An unbalanced signature influences the eigenvalue structure by enforcing a certain number of real eigenvalues and leaving only a few complex ones. As per the theory of matrices, a square matrix with distinct eigenvalues is always diagonalizable ($\det |D| \ne 0$).  Thus the real number of eigenvalues of  $N$, $n \times n$ random matrices will have an interesting statistical distribution. The number of real eigenvalues of a real Gaussian random $n \times n$ matrix is found to be $\sqrt{2n/\pi}$ when $n$ is large, and the real normalized eigenvalue $\epsilon/\sqrt{n}$ of such a random matrix is uniformly distributed over the interval $[-1; 1]$ for large $n$. However, for finite n, distribution \cite{34} of $D({\epsilon})$, real eigenvalues takes the form of an involved analytic function$^{\mbox {\tiny A}}$ of $\epsilon$ and $n$. \footnotetext[1]{\tiny{$D(\epsilon){=}\frac{1}{E_n} \left( \frac{1}{\sqrt{2\pi}} 
			\left[ \frac{\Gamma(n - 1, \epsilon^2)}{\Gamma(n - 1)} \right] 
			+ \frac{|\epsilon|^{n - 1} e^{-\epsilon^2 / 2}}{\Gamma(n/2) 2^{n/2}} 
			\left[ \frac{\gamma((n - 1)/2, \epsilon^2/2)}{\Gamma((n - 1)/2)} \right] 
			\right)$, 
			where, $E_n{=}\frac{1}{2} + \sqrt{\frac{2}{\pi}} \frac{\Gamma(n + 1/2)}{\Gamma(n)} \, {}_2F_1(1, -1/2; n; 1/2)$.}}
	.\\
	\indent Pseudo-symmetric matrices, a form of more general pseudo-Hermitian matrices, with some of the eigenvalues as real can represent PT-symmetric quantum systems having broken PT-symmetry, while pseudo-symmetric matrices with all the eigenvalues as real can represent the systems with exact (unbroken) PT-symmetry \cite{18}, and in more general way, these matrices can made to represent the both the scenario: unbroken PT-symmetry and broken PT-symmetry, under the change of characteristic parameter of the system. In \cite{12}, we have studied the spectral distributions of real eigen-values of the pseudo-symmetric matrices where some eigenvalues are real while others appear as complex conjugate pairs, with $\altmathcal{N} [n(n+1)/2 \le \altmathcal{N} \le n^2]$ Gaussian distributed random numbers as their elements to find the NLSDs as semi-Poisson and sub-Wigner distributions ( intermediate statistics ). Here, we introduce a comprehensive framework of pseudo-symmetric matrices that can exhibit both fully real spectra and mixed spectra with real and complex conjugate pairs. It also captures the transition between these regimes through parameter tuning, providing a unified description of the spectral distributions of PT-symmetric quantum systems in both phases.
	%\twocolumngrid
	%%%%%%%%%%%%%%%%%%%%%%%%%%%%%%%%%%%%%%%%%%%%%%%%%%%%%%%%%%%%%%
	We construct a set of real pseudo-symmetric matrices containing two real parameters $\lambda$ and $\mu$, which are in a hidden way similar and not similar to a real symmetric matrix, and investigate the spectral distributions of the ensemble of pseudo-symmetric matrices using $\altmathcal{ N}=n(n+1)/2$ independent Gaussian\enspace random numbers as their elements, called as Gaussian pseudo-Orthogonal Ensemble (G-pOE) owing the pseudo-orthogonal symmetry of these pseudo-symmetric matrices. We find that, when $\lambda \mu >0$, NLSDs $(p(s))$ come out to be Wigner's surmise as,
	\begin{equation}
		p_W(s)=\frac {\pi}{2}  s \exp \big({-\frac{\pi s^2}{4}}\big)
	\end{equation}
	and distribution of eigenvalues $D(\bar{\epsilon})$ are semi-circle law \cite{1} as,
	\begin{equation}
		D(\bar{\epsilon})=\frac{2}{\pi} \sqrt{1-\bar{\epsilon}^2};~~~ \bar{\epsilon} = \epsilon/\epsilon_{\mbox{max}},
	\end{equation}
	where $\epsilon$ are the eigenvalues of the matrix.
	
	\noindent For $\lambda \mu<0$,  spacing distribution is found to be the intermediate statistics, which fits well to the sub-Wigner form\cite{12,27} given as
	\begin{equation}
		p_{abc}(s){=}a~s~\exp(-b s^c),~0<c<2,
	\end{equation}
	
	\noindent and distribution of eigenvalues $D( \bar{\epsilon})$ for some of these ensemble can be fitted to 
	\begin{equation}
		D(\bar{\epsilon})=A \Big ( \tanh \big(  \frac { \bar{\epsilon}+B}{C} \big )-\tanh \big(  \frac { \bar{\epsilon}-B}{C} \big) \Big ).
	\end{equation} 
	\linebreak 
	\noindent The fitted parameters $a,b,c$;$~$and$~$$A,B,C$ are real and do depend on $n$.$~$Importantly, in complete parametric space of $\lambda$ and $\mu$ ($\lambda,~\mu {~\in~}\rm I\!R_{\ne 0}$), both the type of statistics,  Wigner's surmise and intermediate statistics can be seen to occur.  \\
	The paper is organized as follows: in sec.\enspace II, we construct sets of real pseudo-symmetric matrices and we prove their (hidden) similarity to real symmetric matrices (when $\lambda \mu>0$) and hence the reality of their eigenvalues. The constructed matrices for the case $\lambda \mu <0$ have eigenvalues as both real and complex conjugate pairs. We also discuss the pseudo-orthogonal property for the diagonalizing matrices of these pseudo-symmetric matrices and show that these matrices form the pseudo-orthogonal group. In sec.\enspace IV, we investigate the spectral statistics for the ensemble of constructed pseudo-symmetric random matrices followed by a description of the unfolding procedure in sec.\enspace III. In sec.\enspace V, we derive the NLSD ($p(s)$) for an ensemble of $2 \times 2$ pseudo-symmetric matrices discussed in sec.\enspace II. Finally, we consider the more general form of constructed pseudo-symmetric matrices to find Wigner's distribution and intermediate statistics and then we conclude the present work.
	%%\onecolumngrid
	\vspace{-0.5cm}
	\section{\label{sec:level2} \large Pseudo-Orthogonal Group of New Real Pseudo-symmetric matrices}
	Let $M$ be a real symmetric square matrix of dimension $n$. Let us define the $n \times n$ ($n$ even) Pauli like block matrices $\Sigma_k$ using $n/2 \times n/2$ identity matrices $I$ and $n/2 \times n/2$ null matrices $O$,

	\onecolumngrid
	\begin{eqnarray}
		\Sigma_1(\lambda)=\left (\begin{array}{cc} O & \lambda I\\   I & O \end{array}\right), \quad \Sigma_2(\lambda)= \left (\begin{array} {cc} O & -i \lambda I \\ i I & O \end{array}\right),  \nonumber ~~~\Sigma_3(\lambda)=\left (\begin{array}{cc} \lambda I & O \\ O & - I \end{array}\right),
	\end{eqnarray}
	where\enspace$\lambda{~\in~}\rm I\!R_{\ne 0}$.\enspace Now let us construct the set of matrices $Q_k(\lambda) \!=\!\Sigma_k(\lambda) M \Sigma_k(\lambda)$  and $R_k(\lambda)=\Sigma_k(\lambda) M \Sigma_k^{-1}(\lambda)$, where $k =1,2,3 $. The more generalized form of these matrices, ${\cal Q}_k(\lambda, \mu)=\Sigma_k(\lambda) M \Sigma_k(\mu)$ and ${\cal R}_k(\lambda, \mu)=\Sigma_k(\lambda) M \Sigma_k^{-1}(\mu)$ for $k=1,2,3$ are also constructed. The matrices $Q_k(\lambda)$ and $R_k(\lambda)$ are pseudo-symmetric under the constant metric $\eta_1$ and $\eta_2$ respectively for all $\lambda(\ne 1)$ and for $\lambda=1$, these matrices turn to be the real symmetric matrices. Similarly, the generalized matrices ${\cal Q}_k(\lambda, \mu)$ and ${\cal R}_k(\lambda, \mu)$ are pseudo-symmetric under the constant metric $\eta_3$ and $\eta_4$ respectively for all $\lambda, \mu ~(\lambda=\mu \ne 1)$.  The metrics $\eta_n$ are given as,
	\begin{multline}
		~~~~~~~~~~~~~~\eta_1{=}\left (\begin{array}{cc} \frac{1}{\lambda} I & 0\\   0 & \lambda I \end{array}\right), \enspace  \eta_2{=}\left (\begin{array}{cc} \frac{1}{\lambda^2} I & 0\\   0 & I \end{array}\right), \enspace \eta_3{=}\left (\begin{array}{cc}  \frac{1}{\lambda} I & 0\\   0 &  \mu I \end{array}\right), \enspace  \eta_4{=}\left (\begin{array}{cc} \frac{1}{\lambda \mu} I & 0\\   0 & I \end{array}\right).~~~~~~~~~~~~~~~
	\end{multline}
	Since the matrices $Q_k(\lambda)$ and $ {\cal Q}_k(\lambda, \mu)$ turn out to be real symmetric matrices for the case $k=3$ for all $\lambda, \mu$, so we have not included in our investigation. For brevity, we denote the pseudo-symmetric matrices $Q_k(\lambda)$, $R_k(\lambda)$, ${\cal Q}_k(\lambda, \mu)$ and ${\cal R}_k(\lambda, \mu)$ as $Q_k$, $R_k$, ${\cal Q}_k$ and ${\cal R}_k$ in rest of the paper unless stated otherwise. In the following subsections (A–D), we discuss the pseudo-orthogonal group formed by the diagonalizing matrices, along with general considerations pertaining to pseudo-symmetric matrices.
	\noindent
	\subsection{\label{sec:level2} PSEUDO-SYMMETRY:} Pseudo-symmetry of these real non-symmetric matrices\enspace under constant metrics $\eta_n$ can be seen as follows,
	\begin{multline}
		\hspace{1.4cm} \eta_1 Q_k \eta^{-1}_1=\eta_1 \Sigma_k (\lambda) M \Sigma_k (\lambda) \eta^{-1}_1=\Sigma^t_k(\lambda) M \Sigma^t_k(\lambda)=\Sigma^t_k(\lambda) M^t \Sigma^t_k(\lambda)=Q^t_k \\
		\hspace{-0.1cm} \eta_2 R_k \eta^{-1}_2=\eta_2 \Sigma_k(\lambda) M \Sigma^{-1}_k(\lambda) \eta^{-1}_2=(\Sigma^{-1}_k(\lambda))^t M \Sigma^t_k (\lambda)=(\Sigma^{-1}_k(\lambda))^t M^t \Sigma^t_k(\lambda) =R^t_k \\
		\hspace{-3.8cm} \eta_3 {\cal Q}_k\eta^{-1}_3=\eta_3\Sigma_k(\lambda)M\Sigma_k(\mu)\eta^{-1}_3=(\Sigma_k(\mu))^tM (\Sigma_k(\lambda))^t={ {\cal Q}}^t_k\\
		\hspace*{-34.3cm} \hspace*{-500.3cm} \eta_4 {\cal R}_k \eta^{-1}_4=\eta_4 \Sigma_k(\lambda)M\Sigma^{-1}_k(\mu) \eta^{-1}_4=(\Sigma^{-1}_k(\mu))^t M \Sigma^t_k(\lambda)=(\Sigma^{-1}_k(\mu))^t M^t \Sigma^t_k(\lambda)={ {\cal R}}^t_k. 
	\end{multline}
	Here, the above-mentioned proof of pseudo-symmetry for $Q_k$ and ${\cal Q}_k$ holds  only for   $k=1,2$.
	\subsection{\label{sec:level2} REAL SPECTRUM:} Reality of the spectrum of these real pseudo-symmetric (non-symmetric) can be proved as follows, \\
	
	\noindent{\bf (i)}~$Q_k=\Sigma_k(\lambda) M \Sigma_k(\lambda)=\Sigma^{-1}_k(\lambda) (\Sigma_k(\lambda) \Sigma_k(\lambda) M) \Sigma_k(\lambda){\Rightarrow}~Q_k$ and $\Sigma_k \Sigma_k M$ are\enspace similar\enspace matrices. \\
	\enspace So eig($Q_k$)=eig($\Sigma_k \Sigma_k M$)= $\lambda$~eig($M$). Hence $Q_k(k=1,2)$ matrices will have all the eigenvalues as real.\\
	
	\noindent{\bf(ii)} $R_k=\Sigma_k(\lambda) M \Sigma_k^{-1}(\lambda)=\Sigma_k(\lambda) (M) \Sigma^{-1}_k(\lambda){\Rightarrow}~R_k$ and $M$ are similar matrices.  So eig($R_k$)=eig($M$). Hence $R_k(k=1,2)$ matrices will have all the eigenvalues as real.\\
	
	{\color{black} \noindent{\bf(iii)}~$ { \cal Q}_k(\lambda,\mu) =\Sigma_k(\lambda)M\Sigma_k(\mu)={\Sigma^{-1}_k}{(\mu)}[{\Sigma_k}(\mu){\Sigma_k}(\lambda) M]\Sigma_k(\mu), {\Rightarrow}$~so the matrices ${ \cal Q}_k$ are similar to $\Sigma_k(\mu)\Sigma_k(\lambda) M$.\enspace For $\lambda \mu{>}0$, we can write, $\Sigma_k(\mu)\Sigma_k(\lambda)M =$sgn$(\mu) J_k^2(\mu, \lambda)M =$sgn$(\mu) J_k(\mu, \lambda) [J_k(\mu, \lambda) \\M J_k(\mu, \lambda)] J_k^{-1}(\mu, \lambda)$, where  $J_k^2(\mu, \lambda)=\Sigma_k(\mu)\Sigma_k(\lambda)$. So the pseudo-symmetric matrices $\altmathcal{Q}_k$ are similar to the real symmetric matrices $J_k(\mu, \lambda)M J_k(\mu, \lambda)$ for $\lambda \mu>0$. Hence $\altmathcal{Q}_k(\lambda, \mu)$ will have all real eigenvalues for $\lambda \mu>0$ and partially real for $\lambda \mu<0$. \\
		
		\noindent{\bf(iv)~}$\altmathcal{R}_k(\lambda, \mu){=}\Sigma_k(\lambda) M  \Sigma^{-1}_k (\mu)=\Sigma_k(\lambda)[M \Sigma^{-1}_k (\mu) \Sigma_k (\lambda)] \Sigma^{-1}_k(\lambda)$.$~$For $\lambda \mu$~$>$~$0$, we can write,\enspace$M \Sigma^{-1}_k (\mu) \Sigma_k (\lambda){=}$sgn$(\mu) M  K_k^2(\mu, \lambda){=}$sgn$(\mu) K^{-1}_k(\mu, \lambda) [K_k(\mu, \lambda)M K_k(\mu, \lambda)] K_k(\mu, \lambda)$ where, $K_k^2(\mu, \lambda)=\Sigma^{-1}_k(\mu)\Sigma_k(\lambda)$. So finally the matrices ${\cal R}_k$ are similar to the symmetric matrices $K_k(\mu, \lambda)M K_k(\mu, \lambda)$ for $\lambda \mu>0$. Hence ${\cal R}_k(\lambda, \mu)$ will  display both type of scenarios: all and some of the eigenvalues as real for $\lambda \mu>0$ and $\lambda \mu <0$ respectively.}\\
	
\noindent	It is interesting to note that $\mbox{det}({\cal Q}_k(\lambda, \mu)-x\mbox{\textbf{1}}) = \mbox{det}(\mu {\cal R}_k(\lambda, \mu)-x\mbox{\textbf{1}})$, which implies the eigenvalues of ${\cal Q}_k(\lambda, \mu)$ are $\mu$ times the eigenvalues of ${\cal R}_k(\lambda, \mu)$ for $k=1, 2$. Furthermore, for k = 3, the matrices ${\cal Q}_3(\lambda, \mu)$ and ${\cal R}_3(\lambda, \mu)$ share the same set of eigenvalues.
	\\
	\twocolumngrid
	\subsection{\label{sec:level2} PSEUDO-ORTHOGONALITY:} A real square matrix $A$ is said to be pseudo-orthogonal under a metric $\zeta$, if
	\begin{equation}
		A^t \zeta A =\zeta.
	\end{equation} 
	Let $A$ be $G^{-1}BG$, where $G$ is a square matrix, not necessarily real and $B$ is an orthogonal matrix such that $BB^t=I$. Eq.(7) follows,
	\begin{equation}
		A^t \zeta A = G^t B^t (G^{-1})^t \zeta G^{-1} B G
	\end{equation} 
	Let $(G^{-1})^t \zeta G^{-1} =I \Longrightarrow \zeta= G^tG$. 
	\begin{equation}
	\hspace{-1.8cm}	A^t \zeta A = G^t B^t B G = G^t G = \zeta
	\end{equation}
	Hence, the matrix $A (=G^{-1}BG)$ is pseudo-orthogonal under the metric $\zeta= G^tG$ as $A^t \zeta A =\zeta$.
	\\
	\noindent \textbf{(i)} Let the diagonalizing matrix of $M$ be $D$ as $D^t M D= E$, where E is the diagonal matrix. Since $M$ is a real symmetric matrix ($M=M^t$), so the diagonalizing matrix $D$ is orthogonal as $DD^t=I$. We can find the diagonalizing matrix of $Q_k$ as,
	\begin{multline}
		\qquad \qquad Q_k{=}\Sigma_k M \Sigma_k=\Sigma^{-1}_k (\Sigma_k \Sigma_k M) \Sigma_k\\
		\Sigma_k Q_k \Sigma^{-1}_k = \lambda M\\
		D^{-1} \Sigma_k Q_k \Sigma^{-1}_k D = \lambda E\\
	\hspace{-0.4cm}	(\Sigma^{-1}_k D \Sigma_k)^{-1} Q_k (\Sigma^{-1}_k D \Sigma_k){=}\lambda \Sigma^{-1}_k E \Sigma_k= \lambda E' 
	\end{multline}
	The diagonal matrices $E$ and $E'$ are the same except for the arrangement of elements along the diagonal. Since eig($Q_k$)=eig($\lambda M$), so the matrices ${\cal D}_k=\Sigma^{-1}_k D \Sigma_k$ are the diagonalizing matrices for the pseudo-symmetric matrices $Q_k$. Hence, the non-symmetric matrices ${\cal D}_k $ are  pseudo-orthogonal under the constant metrics $\zeta_k= \Sigma_k \Sigma_k^t$. Here, the metrics $\zeta_k$ turn out to be the same as $\eta_1$, so the diagonalizing matrix for the matrix $Q_k$, which is pseudo-symmetric under the constant metrix $\eta_1$, is pseudo-orthogonal under the same constant metric $\eta_1$.\\
	
	\noindent\textbf{(ii)} Now, on reconsidering the Eq.(7) for a real square matrix $A$= $GBG^{-1}$,
	\begin{equation}
		A^t \zeta A = (G^{-1})^t B^t G^t \zeta G B G^{-1}
	\end{equation} 
	Let $G^t \zeta G =I \Longrightarrow \zeta= (GG^t)^{-1}$. 
	\begin{equation}
		A^t \zeta A = (G^{-1})^t B^t B G^{-1} = (G^{-1})^t G^{-1} = \zeta
	\end{equation}
	Hence, the matrix $A (=GBG^{-1})$ is pseudo-orthogonal under the metric $\zeta= (GG^t)^{-1}$ as $A^t \zeta A =\zeta$. \\
	We can find the diagonalizing matrix of pseudo-symmetric matrices $R_k$ as,
	\begin{multline}
		R_k=\Sigma_k M \Sigma_k^{-1} \Longrightarrow D^{-1} \Sigma^{-1}_k R_k \Sigma_k D = E\\
		(\Sigma_k D \Sigma^{-1}_k)^{-1} R_k (\Sigma_k D \Sigma^{-1}_k){=} \Sigma_k E \Sigma^{-1}_k{=}E''
	\end{multline}
	Again, the diagonal matrices $E$ and $E''$ are the same except for the arrangement of elements along the diagonal. Since eig($R_k$)=eig(M), so the matrices ${\cal D}_k=\Sigma_k D \Sigma^{-1}_k$ are the diagonalizing matrices for the pseudo-symmetric matrices $R_k$. Hence, the matrices $R_k (\ne R^{t}_k)$ are pseudo-orthogonal under the constant metrics $\zeta_k= (\Sigma_k \Sigma_k^t)^{-1}$. The metrics $\zeta_k$ turn out to be same as $\eta_2$. Similarly, the diagonalizing matrices of general pseudo-symmetric matrices ${\cal Q}_k(\lambda, \mu)$
	and ${\cal R}_k(\lambda, \mu)$ can be found as $\Sigma^{-1}_k(\lambda) D \Sigma(\mu)_k$ and $\Sigma(\lambda)_k D \Sigma^{-1}_k(\mu)$ respectively, these in turn are pseudo-orthogonal under constant metrics $\Sigma_k(\lambda) (\Sigma_k(\mu))^t$ and $(\Sigma_k(\lambda) (\Sigma_k(\mu))^t)^{-1}$ respectively.
	\onecolumngrid
		\vspace{-0.2cm}
	\subsection{\label{sec:level2} PSEUDO-ORTHOGONAL GROUP:} On re-writing the condition for pseudo-orthogonality (7) as $\zeta^{-1} A^t \zeta= A^{-1}\Longrightarrow A^\#=A^{-1}$, where the symbol $'\#'$ denotes distortion from orthogonality. We consider the set \({\cal O}\) of all matrices of the form ${\cal D}_k = \Sigma_k^{-1} D \Sigma_k $ , as introduced in the previous section. These matrices satisfy the condition $\altmathcal{D}^t \zeta_1 \altmathcal{D}= \zeta_1$,  where $\zeta_1$ is a fixed pseudo-metric (as discussed in Sec. II.C),  which is found to coincide with $\eta_1$. We define \({\cal O}\) as the set of such matrices \({\cal D}_k\), which forms a pseudo-orthogonal group under matrix multiplication. The group structure with respect to fixed metric $\zeta_1 (=\eta_1)$ is justified as follows\cite{35}: \\
	\noindent ({\bf i}) Pseudo-orthogonal matrices ${\cal D}_k$ are closed under multiplication. Let ${\cal D}_k$ and ${\cal D}_k$ be two arbitrary pseudo-orthogonal matrices. Then,
	\begin{multline*}
		\hspace*{-0.5cm} {({\cal D}_k {\cal D}_m )^\#}{=} \zeta^{-1}_1 ({\cal D}_k {\cal D}_m)^t \zeta_1{=}\zeta^{-1}_1 (\Sigma_m)^t D^t (\Sigma^{-1}_m)^t(\Sigma_k)^t D^t (\Sigma^{-1}_k)^t \zeta_1{=}+\Sigma^{-1}_m D^t \Sigma_m  \Sigma^{-1}_k D^t \Sigma_k{=}({\cal D}_k {\cal D}_m )^{-1}
	\end{multline*}
	({\bf ii}) If ${\cal D}_k$ is pseudo-orthogonal under $\zeta_1$ , then ${\cal D}^{-1}_k$ is also pseudo-orthogonal under the same metric as, 
	\begin{multline*}
		\hspace*{4.0cm} ({\cal D}^{-1}_k)^\#{=}\zeta^{-1}_1({\cal D}^{-1}_k)^t\zeta_1{=}\zeta^{-1}_1\Sigma^{-1}_k(D^t)^{-1} \Sigma_k\zeta_1{=}\Sigma_k D^t \Sigma^{-1}_k{=}{\cal D}_k ~~~~~~~~~~~
	\end{multline*}
	\twocolumngrid
	\noindent({\bf iii}) Pseudo-orthogonal matrices ${\cal D}_k$ are associative under multiplication, and associativity of the arbitrary pseudo-orthogonal matrices  ${\cal D}_k$,  ${\cal D}_m$, and  ${\cal D}_o$  can be verified trivially as $({\cal D}_k ({\cal D}_m {\cal D}_o))^\#{=}({\cal D}_k ({\cal D}_m {\cal D}_o))^{-1}$.  \\
	%\begin{multline}
	%({\cal D}_k ({\cal D}_m {\cal D}_o))^\#{=}\eta^{-1}_2 ({\cal D}_k ({\cal D}_m {\cal D}_o))^t \eta_2{=}...{=}({\cal D}_k ({\cal D}_m %{\cal D}_o))^{-1}
	%\end{multline}
	({\bf iv}) The identity matrix would act as the unit element of this symmetry transformation group.\\
	This confirms that \({\cal O}\), the set of all such pseudo-orthogonal matrices ${\cal D}_k$, forms a group under matrix multiplication with respect to the fixed metric $\zeta_1$. A similar group structure can be established for other sets of pseudo-orthogonal matrices discussed in subsection II-C.
	
	The mapping $\phi{:}D{\mapsto}\Sigma^{-1} D \Sigma$ is a group homomorphism from a subgroup of a general linear map to the pseudo-orthogonal group with respect to $\zeta_1$, preserving group structure: $\phi(D_1 D_2)=\phi(D_1)\phi(D_2)$. Hence, the group structure is preserved via this homomorphism.

	\noindent \emph{Group automorphism}:  Given that $\Sigma_1^{-1}\Sigma_2$ and $\Sigma_1^{-1}\Sigma_3$ are orthogonal, it follows that the set of matrices of the form $\Sigma_k^{-1} {\cal D}_k \Sigma_k $, where each ${\cal D}_k$ is derived from an orthogonal matrix $D$, are also orthogonal, as it results from a similarity transformation of an orthogonal matrix. Consequently, this construction defines a group automorphism of the orthogonal group, realized through similarity transformation by $\Sigma_1$. As a result, the set of matrices ${\cal D}_k$ forms a compact subgroup of the pseudo-orthogonal group. However, it does not span the full pseudo-orthogonal group, which includes elements associated with indefinite metrics and is generally non-compact.
	
	\section{\large \label{sec:level1} Unfolding the Spectrum to find the spectral distributions} To find universal statistical properties of the system, the eigenvalues of random matrices need to be normalized, also called as unfolding \cite{36}, to separate the average behavior of the non-universal spectral density from the universal spectral fluctuations. Unfolding the spectrum is essentially the local re-normalization of eigenvalues in such a way that their mean density of eigenvalues is equal to unity. In this paper, the unfolded eigenvalues $\boldsymbol{\varepsilon}_i$ are obtained as
	\begin{equation}
		\boldsymbol{\varepsilon}_i = N({\epsilon}_i).
	\end{equation}  
	where $\epsilon_i$ are  the true eigenvalues of the random matrices, and  $N(\epsilon)$ is cumulative mean density defined as
	\begin{equation}
		N(\epsilon) = \int_{-\infty}^{{\epsilon}} \rho({\epsilon}) d{\epsilon}.
	\end{equation}
	where $\rho({\epsilon})$ is the mean normalized density of eigenvalues of the random matrices.  The level-spacing distributions are given by the probability function $p(s)$, where $s=|\boldsymbol{\varepsilon}_{i+1}-\boldsymbol{\varepsilon}_i|$.
	%%%%%%%%%%%%%%%%%%%%%%%%%%%%%%%%%%%%%%%%%%%%%%%%%%%%%%%%%%%%%%%%%
	%%%%%%%%%%%%%%%%%%%%%%%%%%%%%%%%%%%%%%%%%%%%%%%%%%%%%%%%%%%%%%%%%
	\section{\label{sec:level1} \large Spectral Distributions for Gaussian pseudo-Orthogonal Ensemble of $N \times N$ Real Random Matrices  $Q_k(\lambda)$, and $R_k(\lambda)$: Wigner's Surmise} 	
	The non-symmetric matrices  $Q_k(\lambda)$ and $R_k(\lambda)$ constructed in sec.{~II} are pseudo-symmetric matrices under generalized $\eta$ having all the eigenvalues as real, which can represent the systems having exact PT-symmetry \cite{18}.  Here, we propose to investigate the spectral distributions $p(s)$ and $D(\bar{\epsilon})$ of Gaussian pseudo-orthogonal ensemble of real random matrices arising from pseudo-symmetric matrices $Q_k(\lambda)$ and $R_k(\lambda)$ for the parameter $\lambda{\in}\rm I\!R_{\ne 0}$. We do so by considering 5000 sampling of matrices $Q_k=\Sigma_k(\lambda) M \Sigma_k(\lambda)$, and $R_k=\Sigma_k(\lambda) M \Sigma^{-1}_k(\lambda)$ of dimension $100 \times 100$, where the real symmetric square matrix $M$ is having $ {\cal N}=n(n+1)/2$ random numbers under Gaussian probability distribution with zero mean and variance 1. We find the NLSDs ($p(s)$) for the ensemble of these pseudo-symmetric random matrices numerically after unfolding\cite{36} the spectrum (17) and $p(s)$ of unfolded energy levels turns out to be Wigner's surmise $p_W(s)$ as in Eq. (1).
	\begin{figure}[h]
		\centering
		\includegraphics[width=8.5 cm,height=5.cm]{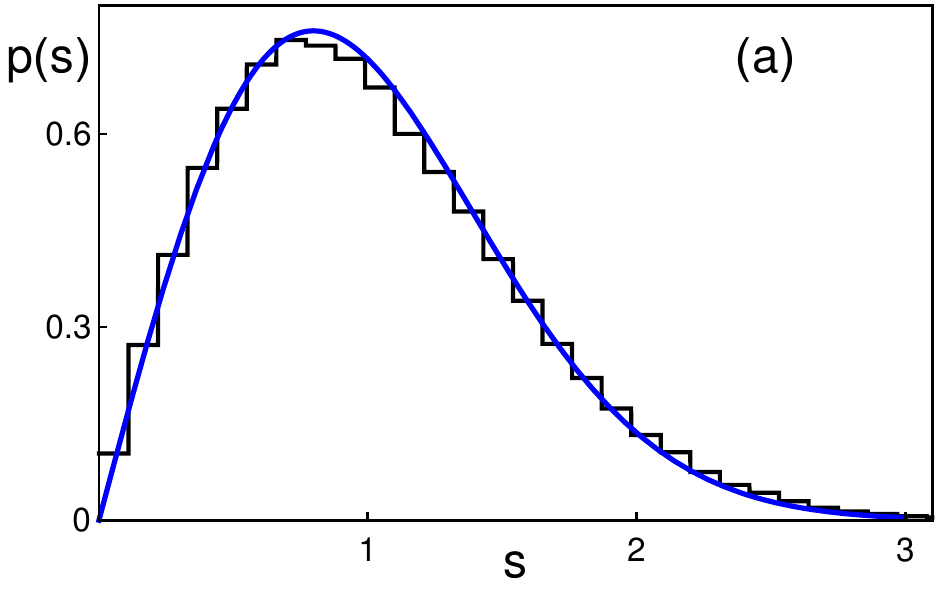}
		\includegraphics[width=8.5 cm,height=5.cm]{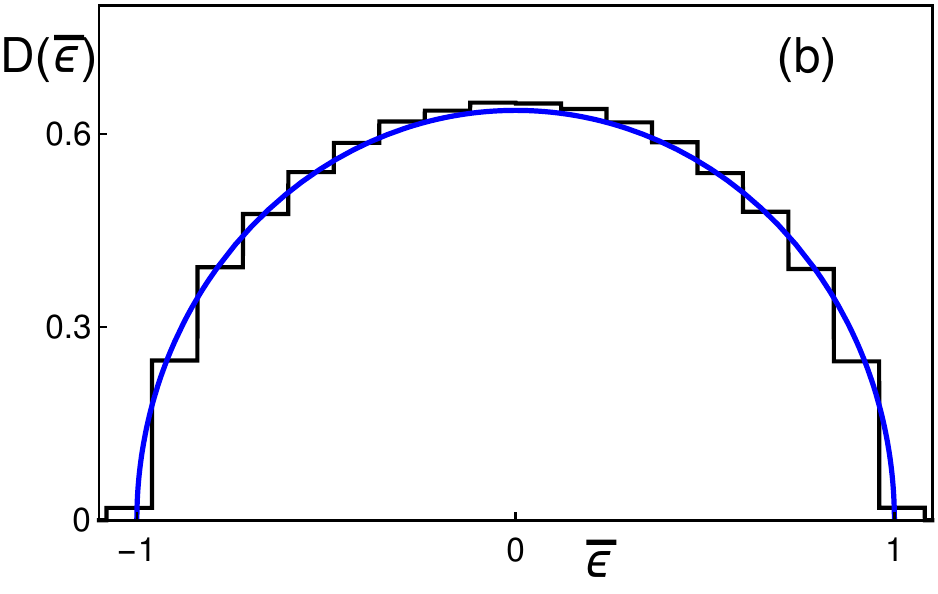}
		\caption{ (a): NLSD histogram for the Gaussion pseudo-orthogonal ensemble (G-pOE) of $5000$, $100 \times 100$ real pseudo-symmetric matrices $Q_1(\lambda)$ for $\lambda=0.5$, plotted against the Wigner's surmise $p_W(s)$, which excellently fits the numerically computed histogram.  These results are insensitive to parameters $\lambda$. (b): Histograms for the average density of eigenvalues $D(\bar{\epsilon})$ for the ensemble of 1000 pseudo-symmetric matrices $Q_1(0.5)$ of order $100 \times 100$ under the Gaussian PDF, plotted against the semi-circle law (2). This is the universality for the cases of other G-pOE of the matrices $Q_2(\lambda)$, $R_{k=1,2,3}(\lambda)$.}
	\end{figure}
	In Fig. 1(a), we have plotted the numerically obtained NLSD histogram $p(s)$ of unfolded energy levels against Wigner's distribution (1) for the ensemble of pseudo-symmetric matrices $Q_1(\lambda)$ for $\lambda =0.5$, the excellence of numerical result with the Wigner's surmise can be seen in Fig 1(a). This demonstrates that spectral statistics are governed by Wigner's surmise, even though the matrices themselves are non-symmetric, broadening the conventional understanding that Wigner's surmise applies only to real symmetric matrices. This extends its applicability beyond real symmetric matrices to PT-symmetric systems with exact PT-symmetry. $~$In Fig.$~$1(b), we have plotted the average density of eigenvalues $D(\bar{\epsilon})$ for the ensemble of pseudo-symmetric matrices $Q_1(0.5)$ along with the semi-circle law (2), the numerically computed histogram matches the semi-circle distribution excellently.   Similar results have also been observed for the ensembles of other pseudo-symmetric random matrices $Q_2(\lambda)$ and $R_{k=1,2,3}(\lambda)$ for all $\lambda{\in}\rm I\!R_{\ne 0}$. Though the distribution of eigenvalues  $D(\bar{\epsilon})$ deviates from semicircle law (2) under the change of parameter $\lambda$, however NLSD after unfolding the spectrum remains invariant as Wigner's surmise under the change of parameter $\lambda$. Ensembles of real pseudo-symmetric matrices with all real eigenvalues having  ${\cal N}=n$ number of matrix elements display the Poisson NLSD \cite{37}. Here, we have considered the pseudo-symmetric matrices with all real eigenvalues having ${\cal N}=n(n+1)/2$, which is very large compared to ${\cal N}=n$. Hence the number ${\cal N}$ of  matrix elements is very crucial in observing Wigner's statistics. \\
	\begin{table}
		%	\large
		\begin{tabular}{|c|c|c|c|c|c|c|c|c|c|}
			\hline
			$\lambda $ & $\mu $ & Real spectrum & $a$ & $b$ & $c$ & Fitted NLSD \\
			\hline
			0.6    & 1.0 & Complete &  $\pi/2$ & $\pi/4$ & 2 & Wigner \\
			\hline
			0.8    & 1.0 & Complete &  $\pi/2$ & $\pi/4$ & 2 & Wigner \\
			\hline
			1.0    &  1.0 & Complete &  $\pi/2$ & $\pi/4$ & 2 & Wigner\\
			\hline	
			-1.0    & -1.0 & Complete &  $\pi/2$ & $\pi/4$ & 2 & Wigner\\
			\hline
			-1.0 & 1.0 & Partial & 6.96 & 2.65 & 0.81 & Sub-Wigner\\
			\hline
			-0.9 & 1.0 & Partial & 6.42 & 2.56 & 0.82 & Sub-Wigner\\
			\hline
			-0.8 & 1.0 & Partial & 4.16 & 2.03 & 1.00 & Sub-Poisson\\
			\hline
			-0.7 & 1.0 & Partial & 2.89 & 1.56 & 1.28 & Sub-Wigner\\
			\hline
			-0.6 & 1.0 & Partial & 2.68 & 1.45 & 1.37 & Sub-Wigner\\
			\hline
		\end{tabular}
		\caption{The parameters of the fitted NLSD function $p_{abc}(s)=a~s~\exp{(-b s^c)}$ (3) to the numerically computed NLSD histograms for the ensembles of pseudo-symmetric matrices ${\cal Q}_k(\lambda, \mu)$ along with corresponding statistics are listed here.}
	\end{table}
	%%%%%%%%%%%%%%%%%%%%%%%%%%%%%%%%%%%%%%%%%%%%%%%%%%%%%%%%%
	%%%%%%%%%%%%%%%%%%%%%%%%%%%%%%%%%%%%%%%%%%%%%%%%%%%%%%%%%%
		\vspace{-0.5cm}
	\section{\label{sec:level1}\large Derivation of NLSD $p(s)$ for Gaussian pseudo-Orthogonal Ensemble of $2 \times 2$ Real Random Matrices}
	\begin{figure}
		%\centering
		\hspace*{-1cm}\includegraphics[width=7.5 cm,height=9cm]{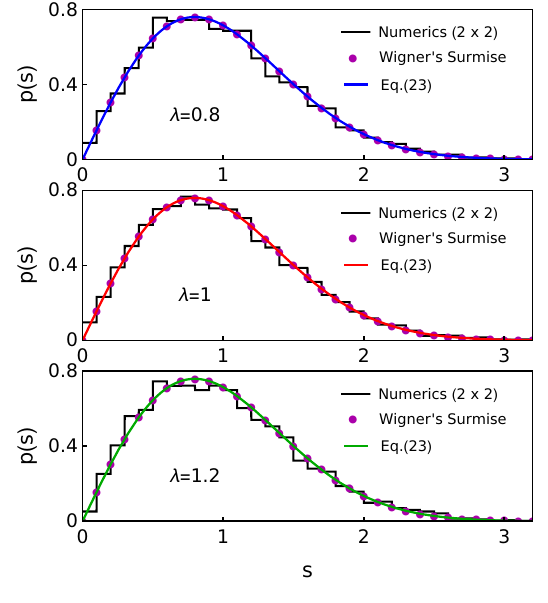}
		\caption{Nearest level-spacing distribution (NLSD) $p(s)$ (21) for the G-pOE of $2 \times 2$ pseudo-symmetric matrices $Q_1(\lambda)$ (16) is shown in Fig. 2 for $\lambda = 0.8$ (blue-solid, upper panel), $\lambda = 1$ (red-solid, middle panel), and $\lambda = 1.2$ (green-solid, lower panel), along with Wigner's surmise $p_W(s)$ (magenta circles). The histogram shows the NLSD for the Gaussian pseudo-orthogonal ensemble (G-pOE) of 5000, real $2 \times 2$ pseudo-symmetric matrices $Q_1(\lambda)$ for the corresponding values of $\lambda$ (black-solid). As is evident, the derived Eq. $p(s)$ (21) converges to $p_W(s)$ and is insensitive to the parameter $\lambda$.}
	\end{figure}
	Let us take a $n=2$ case of the pseudo-symmetric matrix $Q_1(\lambda)= \Sigma_1(\lambda) M  \Sigma_1(\lambda)$ made up of three  $(\altmathcal{ N}= 2(2+1)/2 =3)$ independent elements $a_{11},a_{12},a_{22}$ which comes from a Gaussian-random population, and $\lambda$ is a real fixed parameter. The $Q_1(\lambda)$ is pseudo-symmetric under the constant metric $\eta_1$, 
	\begin{equation}
		Q_1(\lambda){=}\left(\begin{array}{cc} \lambda a_{22} &  \lambda^2 a_{12} \\ a_{12} & \lambda a_{11} \end{array}\right),\quad \hspace{-0.5cm} \eta_1{=}\left (\begin{array}{cc} 1/\lambda & 0 \\  0 & \lambda \end{array}\right),
	\end{equation}
	as $\eta_1 Q_1 \eta_1^{-1}=Q_1^t$. Its eigenvalues $\epsilon_{1,2}=\lambda(a_{11}+a_{22} \mp \frac{1}{2}\sqrt{4a_{12}^2+(a_{11}-a_{22})^2})$ are unconditionally real. The spacing $\epsilon_1-\epsilon_2$ between them is $ \delta \epsilon=\lambda \sqrt{4a_{12}^2+(a_{11}-a_{22})^2}$. As discussed in sec. II, the diagonalising matrix  $\altmathcal{D}_1$ for $Q_1$ is given as $\Sigma^t_1 D \Sigma_1$, which is pseudo-orthogonal under the constant metric $\zeta_1(= \Sigma^t_1 \Sigma_1$) as $\altmathcal{D}^t_1 \zeta_1 \altmathcal{D}_1= \zeta_1$. The orthogonal matrix D and constant metric $\zeta_1$ are given as,	
	\begin{equation}
		D{=}\left (\begin{array}{cc}  \cos \theta & \sin \theta\\  -\sin \theta  &  \cos \theta \end{array}\right), \quad \zeta=\left (\begin{array}{cc}  1 & 0\\  0  &  \lambda^2 \end{array}\right)
	\end{equation}
	where, $\theta= \frac{1}{2} \tan^{-1} \frac{2 a_{12}}{a_{22}{-}a_{11}} \Longrightarrow \theta \in (-\frac{\pi}{4}, \frac{\pi}{4})$. \\ 
	
	\noindent Since $Q_1{=}\altmathcal{D}_1 ( \lambda E') \altmathcal{D}_1^{-1}$ by Eq.(10), so we can write,
	\begin{multline}
		\hspace{1.1cm}a_{11}=\frac{(\epsilon_1+\epsilon_2){-}(\epsilon_1{-}\epsilon_2) \cos(2 \theta)}{2 \lambda}, \\
		a_{12}= \frac{(\epsilon_1-\epsilon_2) \sin(2 \theta)}{2 \lambda}, \\
		a_{22}= \frac{(\epsilon_1+\epsilon_2)+ (\epsilon_1-\epsilon_2) \cos(2 \theta)}{2 \lambda} 
	\end{multline}
	Given a matrix  $Q_1(\lambda)$, the matrix elements are drawn from a Gaussian probability kernel \cite{35,38,21} as
	\begin{multline}
		\hspace{1.5cm} P(Q_1){=}\altmathcal{A} \exp \biggl(-\frac{\mbox{tr}(Q_1 Q_1^{t})}{2 \sigma^2}\biggl) \hspace{.5cm}
	\end{multline}
	\noindent where $\altmathcal{A}$ is the normalization constant, and here $\sigma=1$. Using Eq.(19) and integrating it over $\theta$, we derive the joint probability density function of eigenvalues as,
	\onecolumngrid
	\begin{multline}
		\small
		\hspace{.8cm}P(\epsilon_1,\epsilon_2){=}\altmathcal{ A'}\sqrt{\kappa}(\epsilon_1{-}\epsilon_2)I_0\left (\frac{(\kappa-2) (\epsilon_1-\epsilon_2)^2}{16} \right) \exp \biggl({-\frac{(\kappa+2) (\epsilon_1-\epsilon_2)^2}{16}-\frac{(\epsilon_1+\epsilon_2)^2}{4}} \biggl)
		~~~~~~~%P(E_1,E_2){=} N (E_1-E_2) I_0\left (-\frac{25 (E_1-E_2)^2}{576} \right)e^{-\frac{169 (E_1-E_2)^2}{576}}
	\end{multline}
	where $\kappa{=}\lambda^2+1/\lambda^2$.
	Defining $\epsilon_1-\epsilon_2=\delta\epsilon$ and $\epsilon_1+\epsilon_2=T$, integrating w.r.t. $T$ from -$\infty$ to $\infty$, we get the nearest level-spacing distribution $P(\delta \epsilon)$. Further, defining $s=\delta \epsilon/{<}\delta \epsilon{>}$ and using the normalization as ${<}\delta \epsilon{>}=1$, we find the normalized nearest level-spacing distribution (NLSD) $p(s)$ as 
	%\hspace*{1cm}\fbox{\begin{minipage}{20em}
			\begin{equation}
				%\hspace*{1cm}\fbox{\begin{minipage}{20em}
						p(s,\lambda){=} \frac{\sqrt{2 \kappa}}{\pi} \gamma^2 s~ \exp\biggl(-\frac{(\kappa+2) s^2 \gamma^2}{4 \pi}\biggl) I_0\left (-\frac{(\kappa-2) s^2 \gamma^2}{ 4 \pi} \right)
						%p(s)= A''s e^{-f_1 (\lambda) s^2} I_0 \left( f_2 (\lambda) s^2 \right) 
						%p(s)= \sqrt{\frac{97}{2}} s e^{-\frac{25 %\gamma s^2}{144 \pi}} I_0 \left( \frac{169 %\gamma s^2}{144 \pi} \right) 
					\end{equation}
					
					\noindent where, $\gamma= {\bf E} ((\kappa-2)/\kappa)$ which is the complete elliptic integral\cite{39}, and the derived Eq. (21) is valid for $1/2 \le \lambda \le 3/2$.
					Alternatively, \( P(\delta \epsilon) \) can also be obtained by evaluating the multiple integral involving the delta function \( \boldsymbol{\delta} \),
					\begin{eqnarray}
						P(\delta \epsilon)=\altmathcal{B} \int_{-\infty}^{\infty} da_{11} \int_{-\infty}^{\infty} da_{12}\int_{-\infty}^{\infty} da_{22}~ e^{-(a_{11}^2+a_{12}^2+a_{22}^2)/(2\sigma^2)}   \boldsymbol{\delta}[\delta \epsilon-\lambda \sqrt{4a_{12}^2+(a_{11}-a_{22})^2}], 
					\end{eqnarray}
					which gives,
					\begin{equation}
						P(\delta \epsilon)= \altmathcal{B'} \int_{0}^{\pi/2} \int_{0}^{\pi} e^{-(\delta \epsilon)^2(1+3 \sin^2 \theta))/(4*g^2(\theta, \phi)}) \frac{{(\delta \epsilon)}^2}{|g[\theta, \phi)|^3} \sin \theta ~d\theta~ d\phi,\\
					\end{equation}
					where $g(\theta,\phi)=\lambda \sqrt{1-\sin^2\theta \sin 2 \phi}$.	
					\twocolumngrid		
					\noindent When Eq. (23) is arranged to yield the average spacing as ${<}\delta \epsilon{>}=1$, the normalized (NLSD) $p(s=\delta \epsilon/{<}\delta \epsilon{>})$ converges to $p_W(s)$ (1) for all values of $\lambda$. \\As discussed in sec. II, for $\lambda{=}1$, matrix $Q_1(\lambda)$ becomes a symmetric matrix, hence we recover the expected Wigner's distribution (1) from Eq. (21)\cite{1}. In Fig.$~$2, we have plotted the normalized nearest level-spacing distribution (NLSD) $p(s)$ (21) for different values of $\lambda=0.8$(blue-solid, upper panel), $1.0$(red-solid, middle panel), $1.1$(green-solid, lower panel) along with the $p_W(s)$ (blue-circles). It can be seen that the level spacing distribution $p(s)$ for the $2\times 2$ case of the matrix $Q_1$ remains invariant with respect to the parameter $\lambda$, consistent with the $n\times n$ results discussed in sec. III. Similar results have also been observed for the $n=2$ case of other pseudo-symmetric matrices discussed in sec. II. Several
					other works on ensembles of $2\times2$ pseudo-Hermitian matrices are also worth mentioning here\cite{20}.
					\vspace{-0.5cm}
					\section{\label{sec:level1} \large Spectral Distributions for Gaussian Pseudo-Orthogonal Ensemble of $N \times N$ Real Random Matrices ${\cal Q}_k( \lambda, \mu)$ and ${\cal R}_k( \lambda, \mu)$: Wigner's surmise to intermediate statistics}
					In this section, we consider the general form of pseudo-symmetric matrices $Q_k(\lambda)$ and $R_k(\lambda)$:  ${\cal Q}_k(\lambda, \mu)$ and ${\cal R}_k(\lambda, \mu)$. As discussed in sec. II, for $\lambda \mu > 0$ the eigenvalues of these matrices, are all real, hence their statistics are again Wigner's distribution as found for the pseudo-orthogonal ensemble of matrices $Q_k$ and $R_k$ in sec. III.
					\begin{figure}[h]
						\centering
						\includegraphics[width=8.5 cm,height=5.cm]{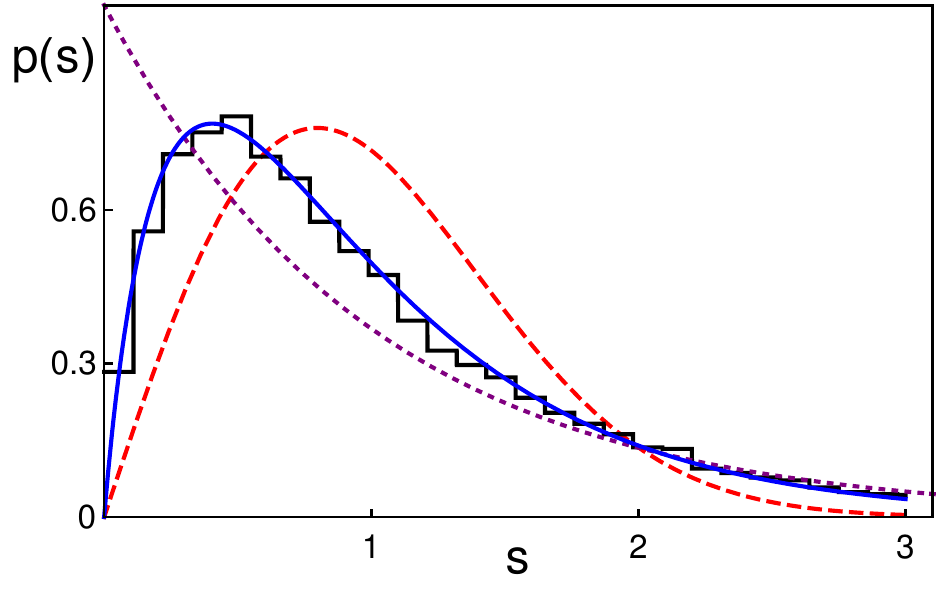}
						\caption{NLSD histogram for G-pOE of $5000$, $100\times100$ real pseudo-symmetric matrices $Q_1(\lambda, \mu)$ for $\lambda=-0.9$, and $\mu=1.0$, plotted against fitted sub-Wigner distributon ($ p(s)=a s e^{-b s^c}, 0 < c < 2$) (blue-solid), alongwith the wigner's surmise $p_W(s)$ (dashed-red) and Poisson-statistics $p_P(s)$ (dotted-purple). See Table I for the parameters $a$, $b$, and $c$, which may change slightly for $n>100$, however, the function form (3) is robust.}
					\end{figure} 
					In Table I, we have listed out the values of the fitted parameter for NLSD ($p(s)=a s e^{-b s^c} $) for some case of $\lambda$ and $\mu$ such that $\lambda \mu > 0$. For $\lambda \mu < 0$, the spectrum of these matrices is partially real, and we find the spectral distributions as intermediate statistics, which are sub-Wigner ($ p(s)=a~s~\exp{(-b~s^c)}, 0{<}c{<}2, c\ne1$) and semi-Poisson ($ p(s)=a~s~\exp{(-b s)}$) depending upon the parameters, as observed in \cite{12}. Notably, these pseudo-symmetric ensembles extend the Gaussian orthogonal ensembles of RMT, leading to the intermediate statistics. In Fig.$~$3, we have plotted the NLSD histogram for the Gaussian pseudo-orthogonal ensemble of 5000, $100 \times 100$ matrices $Q_1(\lambda, \mu)$ for $\lambda=-0.9$ and keeping $\mu=1$, against the fitted NLSD $p(s)=a~s~\exp{(-b~s^c)}$, for the parameters $a=6.96, b=2.65, c=0.81$, which is a sub-Wigner distribution. The distribution of real eigenvalues $D(\bar{\epsilon})$ for most of the ensembles (Table I) of these matrices fits well to the empirical form (4) \cite{12,13} as shown in Fig. 4 for G-pOE of matrices $Q_k(-0.9, 1)$, however $D(\bar{\epsilon})$ deviates from Eq. (4) by sharp rise in number of eigenvalues \cite{13} near $\bar{\epsilon}$=0 as product $\lambda \mu$ increases on negative scale. Remarkably, the analytic result for the distribution\cite{34} of real eigenvalues in $n \times n$ real non-symmetric matrices matches the empirical function $\bar{\epsilon}$ (Eq. 4) in the large-$n$ limit.
					
					\begin{figure}[h]
						\centering
						\includegraphics[width=8.5 cm,height=5.cm]{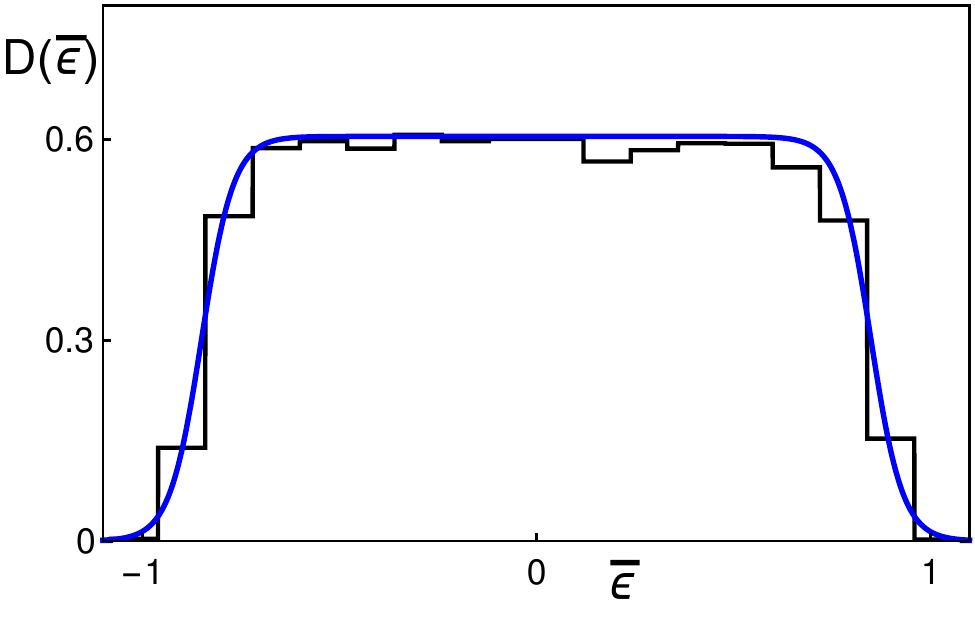}
						\caption{Histograms for the distribution of real eigenvalues $D(\bar{\epsilon})$ for G-pOE of the pseudo-symmetric matrices $Q_1(-0.9, 1.0)$ plotted against fitted empirical function (4). This seems to be the universality for other pseudo symmetric matrices ${\cal Q}_2( \lambda, \mu)$ and ${\cal R}_{k=1,2,3}( \lambda, \mu)$, however, it deviates from Eq. (4) by peaking around $\bar{\epsilon}=0$ as the value of product $\lambda \mu$ increases negatively.}
					\end{figure} 
					
					A well-known PT-symmetric system is a PT-symmetric oscillator\cite{18} with potential $V(x)=(ix)^N$, which exhibits real eigenvalues (unbroken phase) for $N>2$ and complex conjugate pairs (broken phase) for $N\le2$. Similarly, a PT-symmetric quantum optical system\cite{40} with balanced gain and loss terms shows real eigenvalues for small $\gamma$  and complex conjugate pairs for large $\gamma$. Since the pseudo-symmetric matrices ${\cal Q}_k(\lambda, \mu)$ and ${\cal R}_k(\lambda, \mu)$ exhibit this behavior depending on the sign of the product $\lambda \mu$, the spectral distribution of such a class of PT-symmetric systems may be described by the ensembles considered here. When $\lambda \mu>0$, the matrices have purely real eigenvalues, reflecting the unbroken PT-symmetry phase with spectral statistics governed by Wigner's surmise. Conversely, when $\lambda \mu<0$, the spectrum splits into real and complex conjugate pairs, representing the transition to the broken PT-symmetry phase and giving rise to intermediate statistics. The spectral distributions of these matrices may be seen as making the transition from Wigner distribution (1,2) to intermediate statistics (3, 4) as the product $\lambda \mu$ changes from a positive value to negative ($\lambda \mu \ne 0$).$~$Fig.$~$5 shows such a transition from Wigner's distribution to sub-Wigner statistics through semi-Poisson distribution, for the ensemble of matrices ${\cal Q}_1(\lambda, \mu)$ under the change of parameter $\lambda =$ 1.0 (dashed-red), -1.0 (dot-dashed-orange), -0.8 (solid-blue), -0.6 (dotted-black), while $\mu$ is fixed at 1. These features in spectral distribution are also found for the G-pOE of other sets of pseudo-symmetric matrices $Q_2(\lambda, \mu)$ and ${\cal R}_{k=1,2,3}(\lambda, \mu)$.
					\begin{figure}[h]
						\centering
						\includegraphics[width=8.5 cm,height=5.cm]{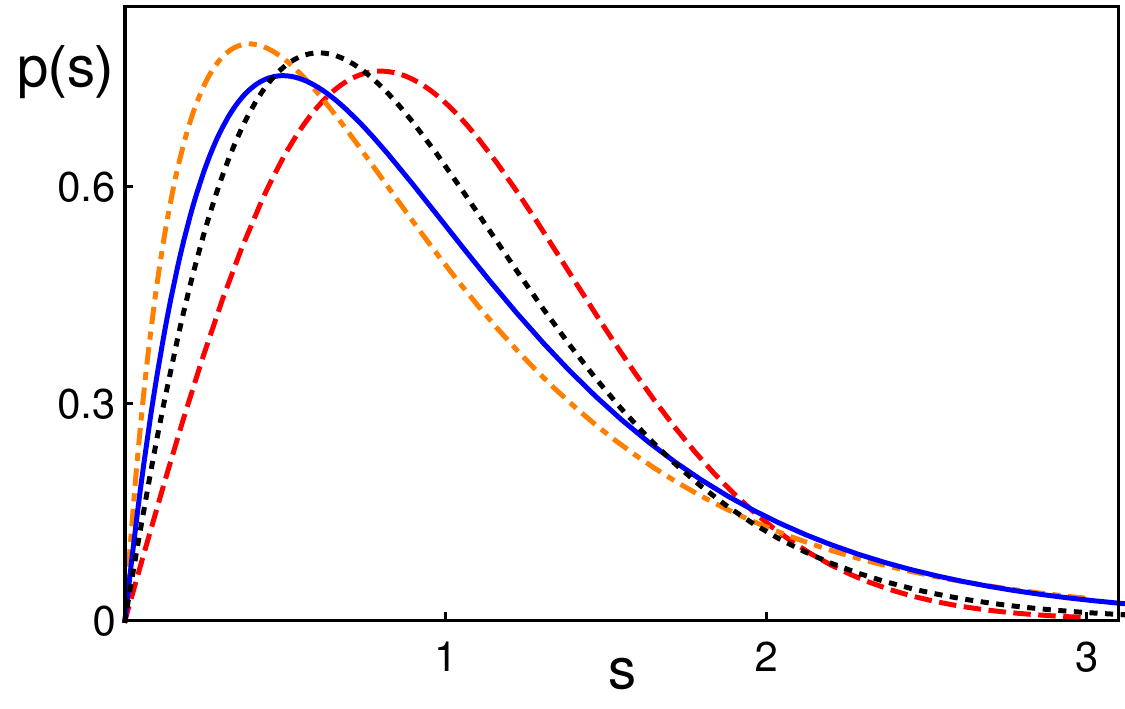}
						\caption{Spacing distribution $p_{abc}(s) = a s\exp(-b s^c)$ obtained by fitting to the numerically computed NLSD histograms for the G-pOE of pseudo-symmetric matrices ${\cal Q}_1(\lambda, \mu)$ under the change of parameter $\lambda=1.0$ (dashed-red), $\lambda=-1.0$ (dot-dashed-orange), $\lambda=-0.8$ (solid-blue), $\lambda =-0.6$ (dotted-black), while $\mu$ is fixed at $1$, presents the transition from Wigner's surmise ($\lambda \mu > 0$) to intermediate statistics ($\lambda \mu < 0$). The corresponding fitted parameters $a$, $b$, and $c$ are listed in Table I. This is the typical universality for other G-pOE of pseudo-symmetric matrices ${\cal Q}_2( \lambda, \mu)$ and ${\cal R}_{k=1,2,3}( \lambda, \mu)$.}
					\end{figure}
					\section{\label{sec:level1} \large Conclusions}
					In this article, the pseudo-symmetric matrices $Q_k(\lambda), R_k(\lambda), {\cal Q}_k(\lambda,\mu; \lambda\mu {>}0)$ and ${\cal R}_k(\lambda,\mu;\lambda\mu>0)$ discussed in sec.\enspace II are new and most interestingly similar to real symmetric matrices in a hidden way and hence their eigenvalues are purely real giving rise to Wigner's surmise yet again.\enspace We claim that the similarity of a pseudo-symmetric matrix to a real matrix is new and thought-provoking. These matrices may be found interesting in general matrix theory as a new type. Here, the Gaussian pseudo-orthogonal ensemble of these random matrices with ${\cal N}= n(n+1)/2$ number of independent Gaussian random numbers has thrown an interesting surprise wherein both the spectral distributions of nearest level spacing and eigenvalues follow Wigner's surmise. This provides the insight that Wigner's surmise is the outcome of matrices whose all eigenvalues are real. These eigenvalues can even unconventionally come from non-symmetric (pseudo-symmetric) matrices as against the conventional real symmetric ones. But when the spectrum splits into separated sets of real and complex conjugate eigenvalues for the pseudo-symmetric matrices ${\cal Q}_k(\lambda,\mu; \lambda\mu{<}0)$ and ${\cal R}_k(\lambda,\mu;\lambda\mu{<}0)$,  spectral distributions display the intermediate statistics.  Since the pseudo symmetric matrices ${\cal Q}_k(\lambda,\mu)$ and ${\cal R}_k(\lambda,\mu)$ can represent the unbroken and broken phase of a PT-symmetric quantum system for $\lambda\mu{>}0$ and $\lambda\mu{<}0$ respectively, thus indicating the connection of un-broken PT-symmetry phase to Wigner's distribution and broken PT-symmetry phase to intermediate statistics. \\
					\noindent More importantly, we have proved that the diagonalizing matrices ${\cal D}$ of these pseudo-symmetric matrices are pseudo-orthogonal under a constant metric $\zeta$ as ${\cal D}^t \zeta {\cal D}=\zeta$, form a pseudo-orthogonal group, and more investigations in this direction are welcome.
					\vspace{-0.5cm}
					\begin{acknowledgments}
						We acknowledge the anonymous referees for their valuable comments, which has significantly improved the manuscript.\\
					\end{acknowledgments}
					
					\section{\label{sec:level1} \large References}
					%\bibliography{refs-rmt}
					%\bibliographystyle{apsrev}
					
				\end{document}